\documentclass[12pt,a4paper]{article}

\usepackage[colorlinks,linkcolor=blue,anchorcolor=red,citecolor=green,urlcolor=blue]{hyperref} 
\usepackage[T1]{fontenc}
\usepackage{amsmath}
\usepackage{amsfonts}
\usepackage{amssymb}
\usepackage{mathrsfs}

\usepackage{graphicx}
\usepackage{color}
\usepackage{soul}
\usepackage{subfig}
\usepackage[normalem]{ulem}
\usepackage[utf8]{inputenc}
\usepackage{authblk}
\usepackage{aas_macros}

\renewcommand{\vec}[1]{\boldsymbol{#1}}

\newcommand{\dif}{\mathrm{d}}

\usepackage{floatrow}
\DeclareFloatFont{tiny}{\scriptsize}
\title{Indication of nearby source signatures of cosmic rays from energy 
spectra and anisotropies}

\begin{document}

\author[a]{Wei Liu}
\author[a]{Yi-Qing Guo\footnote{Corresponding author: guoyq@ihep.ac.cn}}
\author[b,c,d]{Qiang Yuan\footnote{Corresponding author: yuanq@pmo.ac.cn}}
\affil[a]{Key Laboratory of Particle Astrophysics, Institute of High Energy
Physics, Chinese Academy of Sciences, Beijing 100049, China}
\affil[b]{Key Laboratory of Dark Matter and Space Astronomy, Purple Mountain 
Observatory, Chinese Academy of Sciences, Nanjing 210008, China}
\affil[c]{School of Astronomy and Space Science, University of Science and Technology of China, Hefei 230026, China}
\affil[d]{Center for High Energy Physics, Peking University, Beijing 100871, China}

\renewcommand*{\Affilfont}{\small\it} 
\renewcommand\Authands{ and } 

\maketitle

\hfill

\begin{abstract}
The origin of Galactic cosmic rays (GCRs) remains a mystery after more
than one century of their discovery. The diffusive propagation of charged
particles in the turbulent Galactic magnetic field makes us unable to 
trace back to their acceleration sites. Nevertheless, nearby GCR 
source(s) may leave imprints on the locally measured energy spectra
and the anisotropies of the arrival direction. In this work we propose
a simple but natural description of the GCR production and propagation,
within a two-zone disk-halo diffusion scenario together with a nearby
source, to understand the up-to-date precise measurements of the energy
spectra and anisotropies of GCRs. We find that a common energy scale of 
$\sim100$ TeV appears in both energy spectra of protons and helium nuclei 
measured recently by CREAM and large-scale 
anisotropies detected by various experiments. 
These results indicate that one or more local 
sources are very likely important contributors to GCRs below 100 TeV.
This study provides a probe to identify source(s) of GCRs by means of 
joint efforts of spectral and anisotropy measurements.
\end{abstract}


\section{Introduction}

It is widely postulated that GCRs below the so-called knee are mainly 
accelerated by supernova remnants (SNRs), through the well-known diffusive 
shock acceleration process \cite{1978MNRAS.182..147B,1978ApJ...221L..29B}.
A power-law spectrum is expected to be produced at the acceleration source,
i.e., $dN/d{\cal R} \propto {\cal R}^{-\nu}$, with ${\cal R}$ being the
rigidity of the particle. The diffusive transport of GCRs in the 
Milky Way further softens the spectrum by ${\cal R}^{-\delta}$ with 
$\delta\approx0.3\sim0.5$, as suggested by the secondary-to-primary ratio 
of GCRs \cite{2018PhRvL.120b1101A,2017PhRvD..95h3007Y}. This general picture
successfully explains the basic observational properties of GCRs below 
$\sim$PeV, as well as diffuse $\gamma$-rays \cite{2007ARNPS..57..285S}.
However, the GCR anisotropy \cite{1996ApJ...470..501A,2006Sci...314..439A,
2009ApJ...692L.130A,2016ApJ...826..220A,2017ApJ...836..153A} is for a long 
time an unresolved problem. The diffusion model predicts one order of 
magnitude higher of the anisotropies of the arrival directions of GCRs 
compared with the measurements \cite{2012PhRvL.108u1102E}. Meanwhile the 
phase does not point to the Galactic center less than $\sim 100$ TeV as 
expected by the conventional diffusion model \cite{2017ApJ...836..153A}.

Recent precise measurements of the energy spectra of GCRs further
challenge this simple picture, such as the spectral hardenings at 
${\cal R} \sim 200$ GV \cite{2009BRASP..73..564P,2010ApJ...714L..89A,
2011Sci...332...69A,2015PhRvL.114q1103A}, and the spatial variations of 
the inferred energy spectra of GCRs in the Milky Way from Fermi-LAT 
diffuse $\gamma$-rays \cite{2016PhRvD..93l3007Y,2016ApJS..223...26A}. 
These new results suggest in general a non-uniform diffusion scenario 
of GCRs in e.g., the disk and halo \cite{2012ApJ...752L..13T,
2018PhRvD..97f3008G}. This is quite natural that GCRs diffuse slower 
in the Galactic disk where the magnetic field is more turbulent than 
that in the halo. Importantly, it was shown that this two-zone disk-halo 
diffusion scenario can help reduce the predicted amplitude of the GCR
anisotropies \cite{2018PhRvD..97f3008G}. However, it is not a full
solution of the anisotropy problem, since the phase is not satisfactorily
reproduced.

Most recently the balloon-borne experiment CREAM reported new measurements 
of the GCR proton and helium spectra up to $\sim100$ TeV, which revealed 
potential spectral softenings above $\sim20$ TeV \cite{2017ApJ...839....5Y}.
Evidence of similar features was also reported by the NUCLEON 
group \cite{2018JETPL.108....5A}. It is interesting to note that the energy 
distribution of the anisotropy amplitude also becomes flat from $\sim10$ 
TeV and then decreases to a minimum at $\sim100$ TeV after that the 
anisotropy increases again \cite{2017ApJ...836..153A}. The phase of the 
dipole component of the anisotropies changes from R.A.$\sim 4$ hrs around 
$100$ GeV to about $-6$ hrs above $100$ TeV. In particular, the phase 
changes suddenly at $\sim100$ TeV, which implies a paradigm shift at such 
an energy. The common features from $10$ to $100$ TeV of the GCR energy 
spectra and anisotropies suggest a common origin of them.

It has been proposed that the local magnetic field may 
regulate the anisotropies of GCRs due to the anisotropic diffusion, and 
may explain the large-scale anisotropy pattern 
\cite{2010ASTRA...6...49A,2014Sci...343..988S,2015ApJ...809L..23S, 
2016PhRvL.117o1103A,2015PhRvL.114b1101M}. To account for the energy 
dependence of the amplitude and phase of the dipole component of the 
large scale anisotropies, local source(s) may also be necessary 
\cite{2006APh....25..183E,2012JCAP...01..010B,2012JCAP...01..011B,
2013ApJ...766....4P,2013APh....50...33S,2014ApJ...785..129K, 
2015ApJ...809L..23S,2016PhRvL.117o1103A}. Some additional effects, such 
as the motion of the solar system with respect to the local interstellar 
medium and/or the possible limited reconstruction capabilities of 
ground-based experiments are employed to reproduce the observations 
\cite{2016PhRvL.117o1103A}.

In this work, we propose a simple picture, based on the 
spatially dependent propagation (SDP) scenario together with a local source, 
to account for the observational facts about the spectral features of GCRs and 
anisotropies. The SDP model is well-motivated by 
the latest observations on the $\gamma$-ray halos around pulsars by HAWC 
\cite{2017Sci...358..911A}. It has been shown that the SDP scenario can 
also suppress the dipole anisotropies of cosmic rays, and thus help 
reconcile the long-term discrepancy of the anisotropies between data and 
the canonical diffusion model \cite{2018PhRvD..97f3008G}. We suggest that 
new observations of the spectral softenings of the GCR nuclei above 20 TeV 
provide additional support of this scenario.

\section{Model}

\subsection{Spatially-dependent diffusion} 

The shape of the diffusive halo is usually approximated to be a cylinder.
The radial boundary of this propagation halo is equivalent to the Galactic 
radius, i.e., $R = 20$ kpc, whereas its half thickness $z_h$ is about a few
kpc which needs to be determined by fitting the GCR data
\cite{2017PhRvD..95h3007Y,2019SCPMA..6249511Y}. Both GCR sources and the 
interstellar medium (ISM) chiefly spread within the Galactic disk, whose 
width $z_s$ is set to be $\sim200$ pc. Besides the diffusion effect, GCR 
particles may also go through convection, reacceleration, and fragmentation 
due to the collisions with the ISM. At low energies, GCR nuclei further 
lose their energies via the ionization and Coulomb scattering. 
The transport equation is generally written as
\begin{eqnarray}
\frac{\partial \psi}{\partial t} &=& Q(\vec{r}, p) + \nabla \cdot ( D_{xx}\nabla\psi - \vec{V}_{c}\psi )
+ \frac{\partial}{\partial p}\left[p^2D_{pp}\frac{\partial}{\partial p}\frac{\psi}{p^2}\right]
\nonumber\\
&& - \frac{\partial}{\partial p}\left[ \dot{p}\psi - \frac{p}{3}(\nabla\cdot\vec{V}_c)\psi \right]
- \frac{\psi}{\tau_f} - \frac{\psi}{\tau_r} ~,
\label{propagation_equation}
\end{eqnarray}
where $\psi = \dif n/\dif p$ is the CR density per particle momentum $p$ 
at position $\vec{r}$, $Q(\vec{r},p)$ is the source function, $D_{xx}$ and 
$D_{pp}$ are the diffusion coefficients in the space and momentum space
(describing the reacceleration), $\vec{V}_c$ is the convection velocity, 
$\dot{p}$ is the energy loss rate, $\tau_f$ and $\tau_r$ are the fragmentation
and radioactive decaying time scales. At the border of the halo, free escape 
of CRs is assumed, namely $\psi(R, z, p) =  \psi(r, \pm z_h, p) = 0$. 
For a comprehensive introduction to the CR transport, one can refer to 
\cite{2007ARNPS..57..285S,2015ARA&A..53..199G}.

Following \cite{2018PhRvD..97f3008G}, the diffusion coefficient is assumed
to be different in the inner halo ($|z|<\xi z_h$) and outer halo 
($z\ge \xi z_h$), where $\xi \approx 0.1$ characterizes the thickness of 
the disk. In the inner halo region, which is close to the Galactic disk, 
the level of turbulence is appreciably affected by the activities of 
supernova explosions and expected to be intense. Recent HAWC observations 
have shown that the diffusion coefficient of GCRs within tens of parsecs 
around the source is at least two orders of magnitude smaller than the 
conventional one \cite{2017Sci...358..911A}. Since the filling factor of 
such slow diffusion regions is unclear, here we adopt a diffusion 
coefficient in the inner halo in between the HAWC-deduced value and the 
conventional one to approximate an average effect. In the outer halo, 
the turbulence is believed to be CR-driven and less affected by the 
stellar activities. The diffusion coefficient thus reduced to the 
conventional values. The diffusion parameters in the inner and outer
halo are connected smoothly \cite{2018PhRvD..97f3008G}. 
The parameterized diffusion coefficient adopted in this 
work is \cite{2018PhRvD..97f3008G, 2018arXiv180203602L}
\begin{equation}
D_{xx}(r, z, \mathcal R) = D_0 F(r, z) \left(\frac{\mathcal R}{\mathcal R_0} \right)^{\delta_0 F(r, z)} ~.
\end{equation}
$F(r,z)$ is parameterized as
\begin{equation}
 F(r,z) =
\begin{cases}
g(r,z) +\left[1-g(r,z) \right] \left(\dfrac{z}{\xi z_0} \right)^{n} , &  |z| \leqslant \xi z_0 \\
1 , & |z| > \xi z_0
\end{cases} , \\
\end{equation}
in which $g(r,z) = N_m/[1+f(r,z)]$, and $f(r,z)$ is the source density
distribution. The spatial distribution of sources takes the form of SNR 
distribution \cite{1996A&AS..120C.437C}, $f(r,z) \propto 
(r/r_{\odot})^{1.69}\exp[-3.33 (r-r_{\odot})/r_{\odot}]\exp(-|z|/z_s)$, 
where $r_{\odot}=8.5$ kpc and $z_s=0.2$ kpc.
The propagation equation of GCRs is solved with the DRAGON code 
\cite{2008JCAP...10..018E}. The corresponding transport parameters are given 
in Table \ref{tab:transport}. The GCR secondary-to-primary ratios can be 
reasonably reproduced with these parameters \cite{2018arXiv180203602L}.

The injection spectrum of background sources is assumed to be an 
exponential cutoff power-law form of rigidity, $q({\cal R})\propto 
{\cal R}^{-\nu}\exp(-{\cal R}/{\cal R}^\prime_c)$. The cutoff rigidity, 
${\cal R}^\prime_c=6.5$ PV, is tuned to fit the proton and helium spectra 
observed by KASCADE \cite{2013APh....47...54A}. The injection power indexes and normalization fluxes at $E_k = 100$ GeV/n of heavier nuclei refer to \cite{1674-1137-42-7-075103}, which are given in Table \ref{tab:para_inj}.


\subsection{Local source}

The propagation of particles from the local source is calculated using 
the Green's function method, assuming a spherical geometry with 
infinite boundary conditions. The GCR density as a function of space,
rigidity, and time is
\begin{equation}
\phi(r,{\cal R},t)=\frac{q_{\rm inj}({\cal R})}{(\sqrt{2\pi}\sigma)^3}
\exp\left(-\frac{r^2}{2\sigma^2}\right),
\end{equation}
where $q_{\rm inj}({\cal R})\delta(t)\delta({\bf r})$ is the instantaneous 
injection spectrum of a point source, $\sigma({\cal R},t)=\sqrt{2D({\cal R})t}$ 
is the effective diffusion length within time $t$, $D({\cal R})$ is the 
diffusion coefficient which was adopted as the disk value described above. 
The injection spectrum is again parameterized as a cutoff power-law form, 
with a power-law index of $2.20$ ($2.15$) for protons (helium nuclei) and 
a cutoff rigidity of $\sim 60$ TV. 
Note that in this work the local source is assumed to
contribute mainly the proton and helium components of GCRs. The extension
of this work to heavier nuclei can be found in an accompany work
\cite{2019arXiv190512505Q}. The normalization is determined through 
fitting the GCR energy spectra, which results in a total energy of 
$\sim 2\times 10^{50}$ erg for protons and $\sim 1\times10^{50}$ erg for 
helium, which is about $30\%$ of the shock kinetic energy of a typical 
core-collaspe supernova. The distance and age of the local source are 
set to be $d = 330$ pc and $\tau = 3.4 \times 10^5$ years, which are 
the same as that inferred from the observations of Geminga 
\cite{1994A&A...281L..41S,2005AJ....129.1993M,2007Ap&SS.308..225F}.

\section{Results}

\begin{figure}[!ht]
\includegraphics[width=0.48\textwidth]{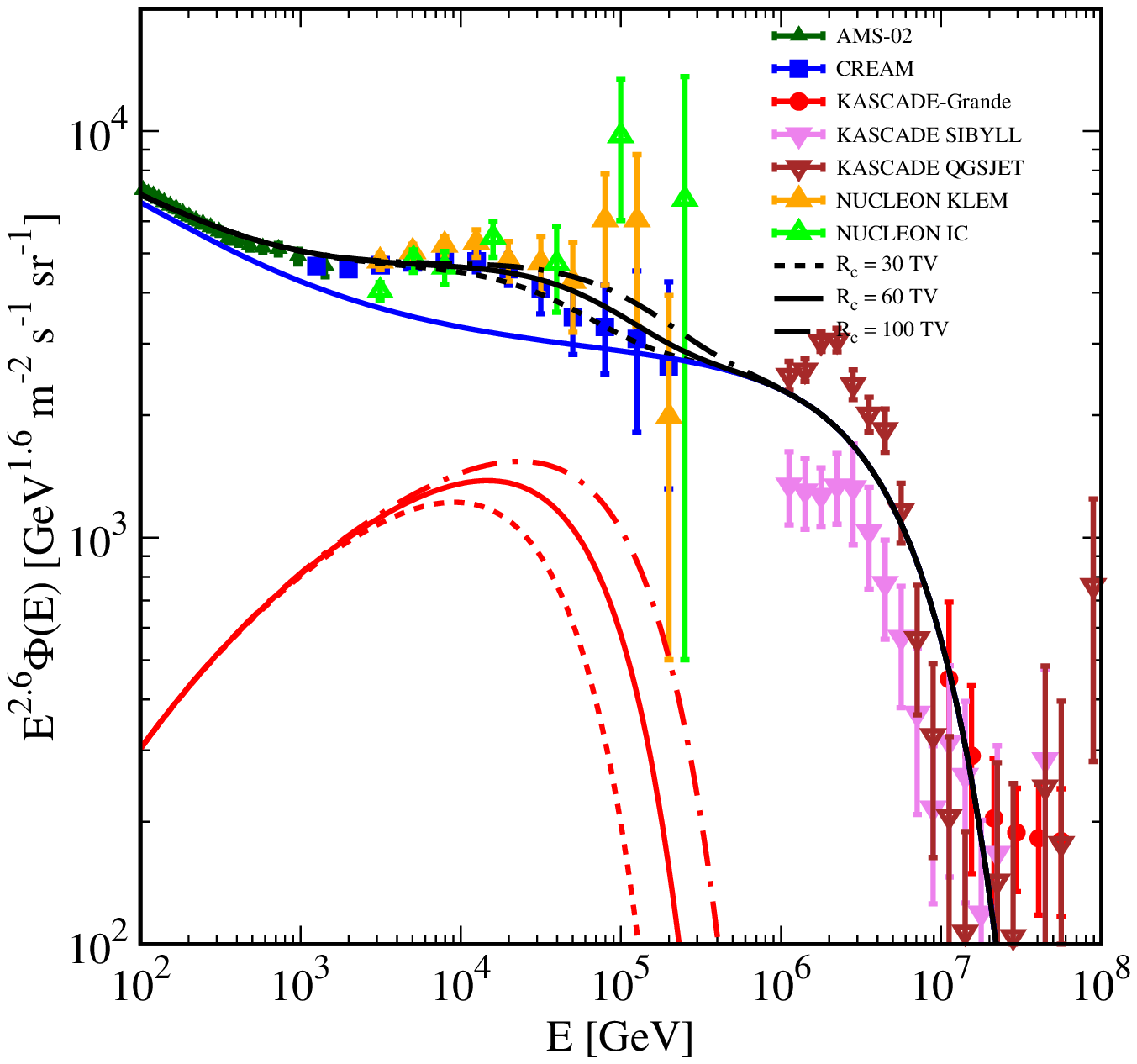}
\includegraphics[width=0.48\textwidth]{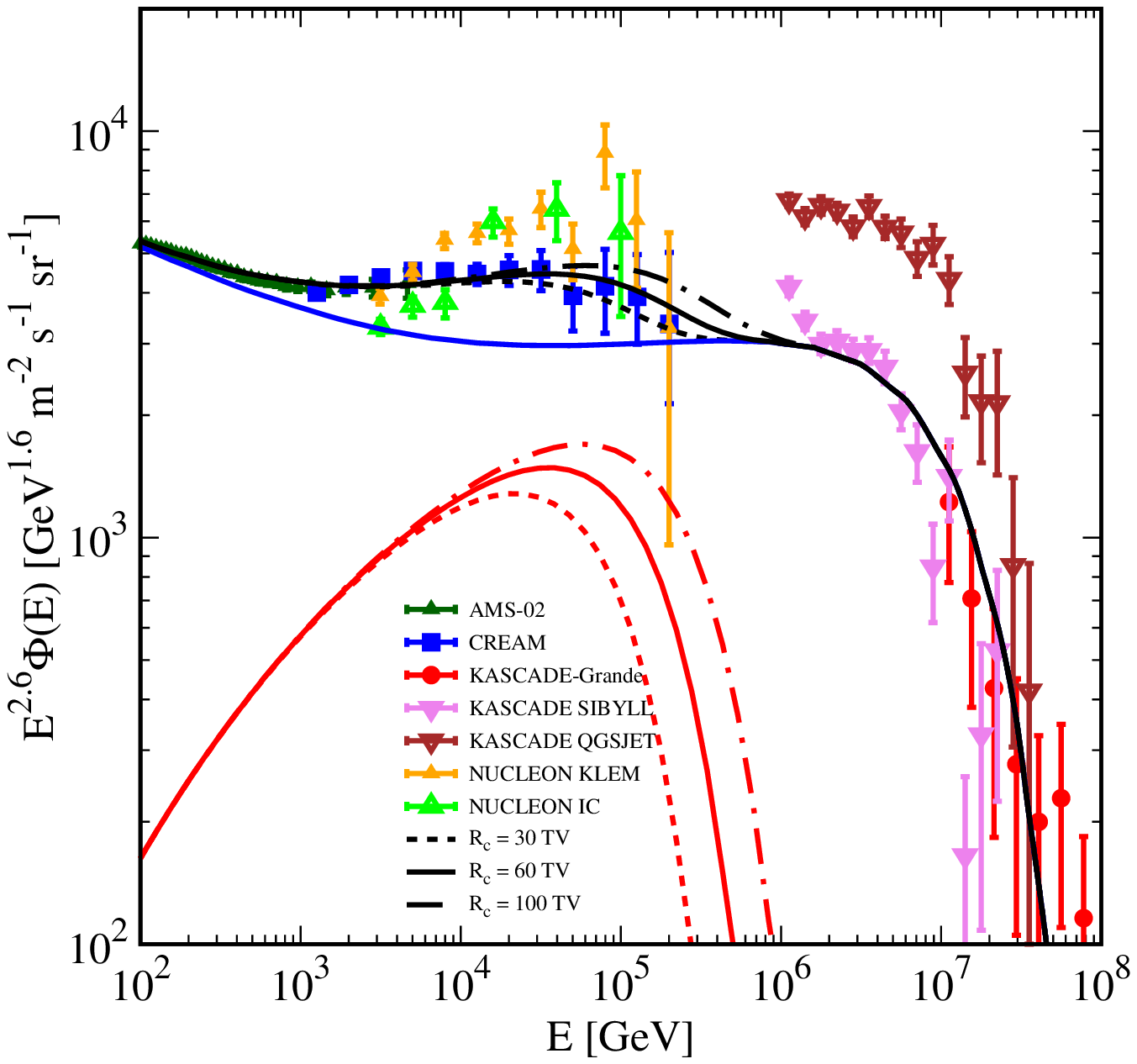}
\caption{Energy spectra of protons (left) and helium nuclei (right). The data 
points are taken from AMS-02 \cite{2015PhRvL.114q1103A,2017PhRvL.119y1101A},
CREAM-III \cite{2017ApJ...839....5Y}, NUCLEON \cite{2017JCAP...07..020A}, KASCADE \cite{2005APh....24....1A} and KASCADE-Grande \cite{2013APh....47...54A} respectively. The blue lines are the background fluxes, the red lines are the fluxes from a nearby source located at (R.A.$=4^h0^m$, $\delta=24^{\circ}30'$) with a distance of $\sim0.3$ kpc, and the black lines are the sum of them. The dotted, solid, and dash-dotted lines are for three different cutoff rigidities, ${\cal R}_c=30$, $60$, and $100$ TV, respectively.
}
\label{fig:Fig1}
\end{figure}

Figure \ref{fig:Fig1} shows the energy spectra of protons and helium from 
the model predictions compared with the measurements by AMS-02 
\cite{2015PhRvL.114q1103A, 2017PhRvL.119y1101A}, 
CREAM-III \cite{2017ApJ...839....5Y}, NUCLEON \cite{2017JCAP...07..020A}, 
KASCADE \cite{2005APh....24....1A} and KASCADE-Grande 
\cite{2013APh....47...54A}. The red, blue and black lines represent the 
contributions from the local source, the background sources and the sum 
of them, respectively. Due to the large measurement errors, 
the value of the cut-off rigidity of the local source contribution, 
${{\cal R}_c}$, has large uncertainties. Here we set ${{\cal R}_c}$ to be
$30$, $60$ and $100$ TV, and find that all of them are consistent with the 
measurements. As we will see below, the anisotropy features are more 
sensitive to the value of ${{\cal R}_c}$. As show in the figure, the background spectrum gradually 
flattens. This is attributed to the SDP effect. The diffusion coefficient 
and its rigidity dependence are assumed to be different in the disk and 
halo regions in the SDP model. Particularly, the diffusion coefficient 
depends more weakly on rigidity in the disk than in the halo. Therefore 
after propagation, the spectrum shows a gradually broken power-law form. 
We find that the recent measurements of the bump-like features of the 
energy spectra of protons and helium by CREAM \cite{2017ApJ...839....5Y} 
and NUCLEON \cite{2018JETPL.108....5A} can be well reproduced in our model. 
Both measurements suggest spectral softenings above tens of TeV, which can 
be a signature of the local source component. 

\begin{table}
\begin{center}
\caption{parameters of SDP model.}
\begin{tabular}{|ccccccccc|}
  \hline
  & $D_0$    &   $\delta_0$     &    $N_{\rm m}$    &    $\xi$   &    $n$  &    $v_A$    &    $z_h$ &  \\
  & [${\rm cm}^2 \cdot {\rm s}^{-1}$] & & & & & [${\rm km}\cdot{\rm s}^{-1}$] & [kpc]  &\\
  \hline
  & $4.46 \times 10^{28}$  & 0.62    & 0.4    & 0.1    & 4    & 6          & 5        &            \\
   \hline
\end{tabular}
\label{tab:transport}
\end{center}
\end{table}

\begin{figure}[!ht]
\includegraphics[width=0.48\textwidth]{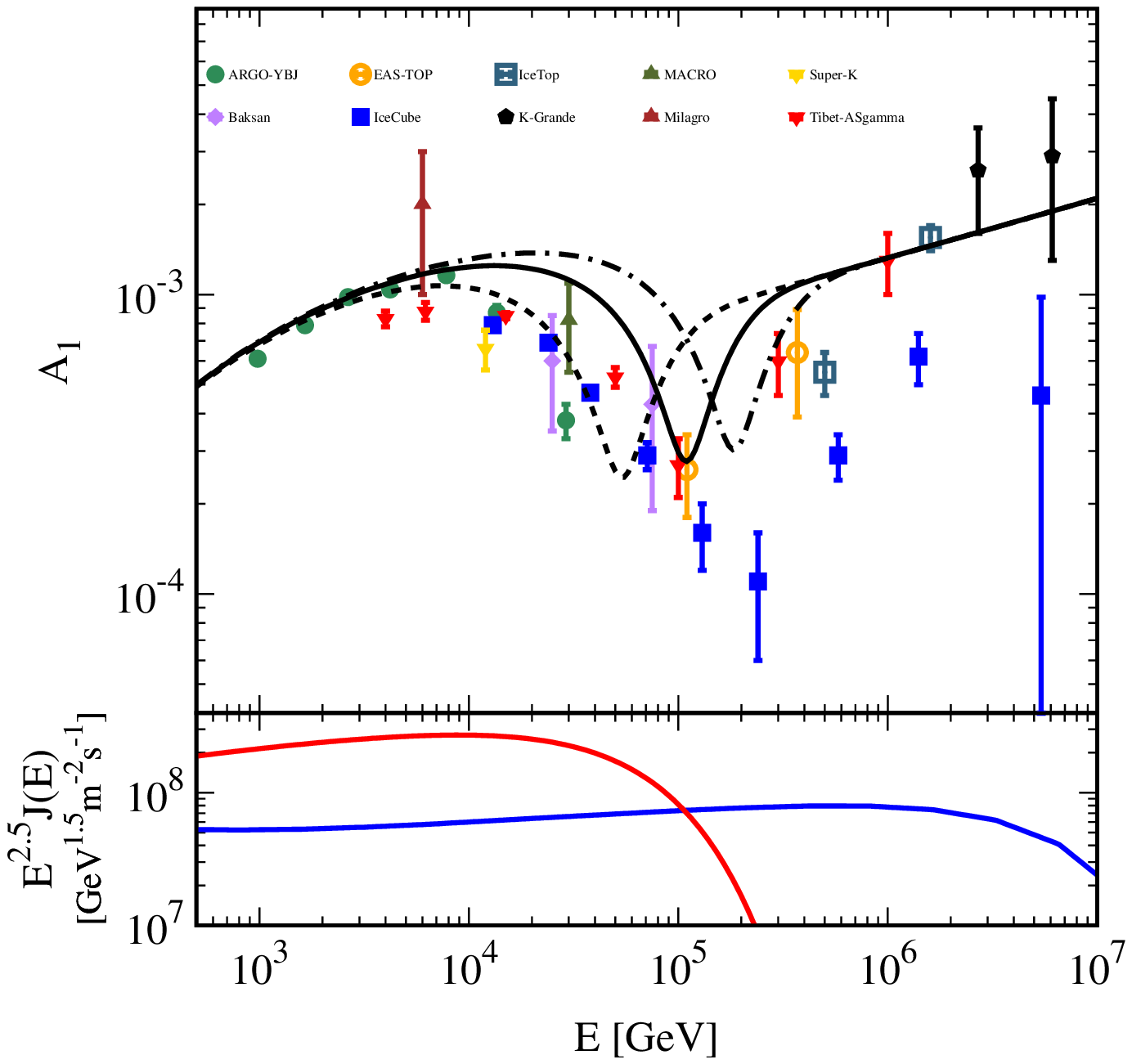}
\includegraphics[width=0.48\textwidth]{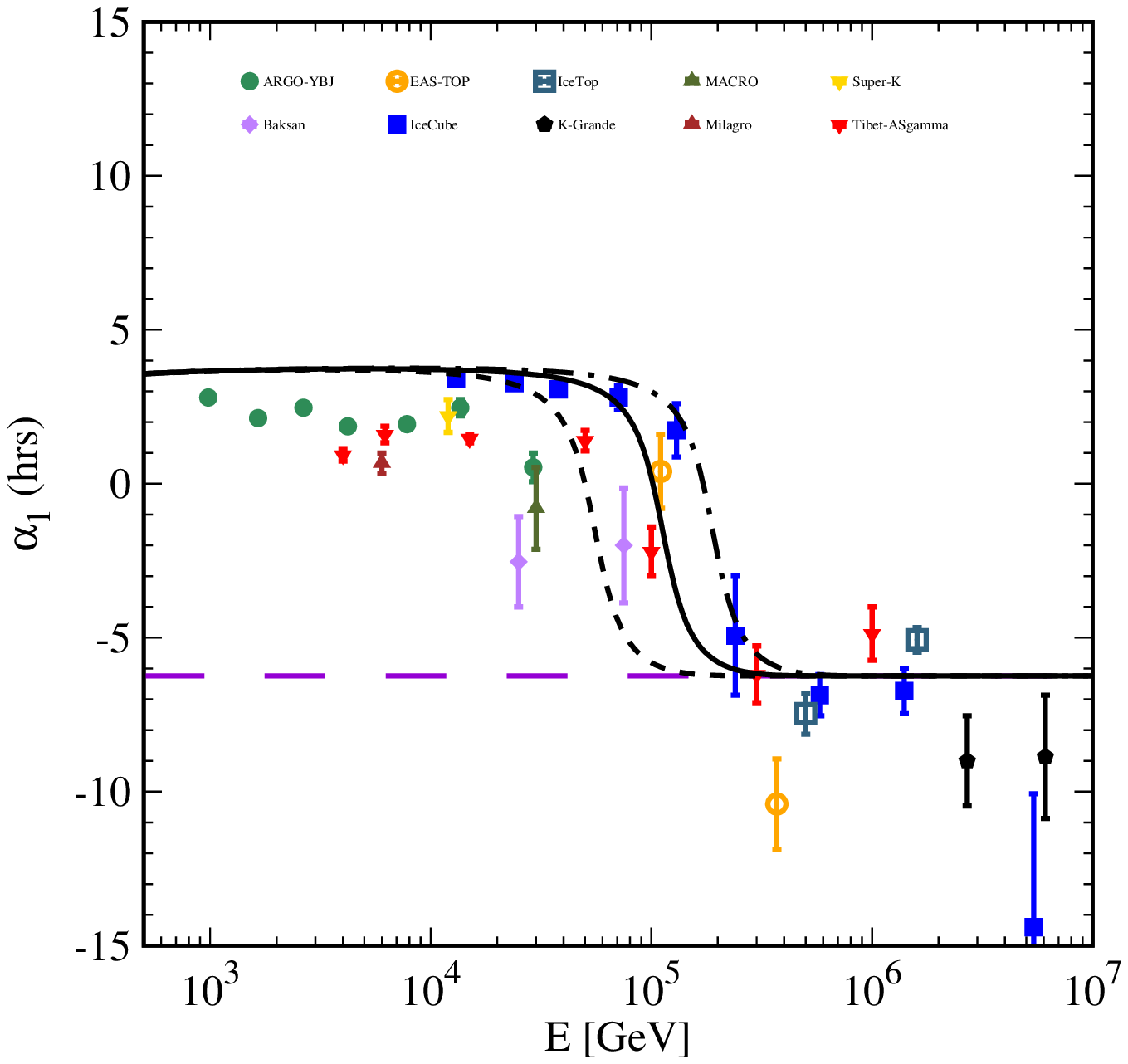}
\caption{Energy dependences of the amplitude (left) and phase (right)
of the dipole anisotropies. In the left panel, the three black lines
correspond to the results for three different cutoff rigidities of the
local source spectra (same as that in Figure~\ref{fig:Fig1}). The blue
and red lines in the lower sub-panel show the flows (defined as $J(E) 
= |D(E) \nabla \phi|$) of the background and local source (for the $60$ TV
cutoff case) components, respectively. In the right panel, the three
black lines represent again the expected phase evolution for the three
cutoff rigidities, and the purple long-dashed line shows the expectation
of the background component, which points from the Galactic center to
the anti-center. Observational data are taken from
ref.~ \cite{2017PrPNP..94..184A} and references therein.
}
\label{fig:Fig2}
\end{figure}

The amplitude and phase of the dipole anisotropy are shown in Figure
\ref{fig:Fig2}. The anisotropy of GCRs depends on the sum of the GCR 
flows from the background (${\bf J}_{\rm bkg}$) and the local source 
(${\bf J}_{\rm local}$). ${\bf J}_{\rm bkg}$ points from the Galactic 
center to the anti-center, since GCR sources are more abundant in the 
inner Galaxy. The direction of the local source can be determined by 
the observational phase of the anisotropy, which suggests that the local 
source is located at the direction of the anti-Galactic center and is out 
of the Galactic disk. We find that a source located at (R.A.$=4^h0^m$, 
$\delta=24^{\circ}30'$) gives very good fit to the measurements of both 
the amplitude and phase of the anisotropy. For $E<100$ TeV, the local 
source contribution dominates the observed anisotropies, although its 
flux is sub-dominant. The phase thus keeps tracing the direction of the 
local source. Meanwhile since the energy spectra of ${\bf J}_{\rm local}$ 
peak around $10$ TeV, the amplitude of anisotropy also peak at such energies. 
For $E\gtrsim100$ TeV, the contribution from the local source decreases 
significantly, and ${\bf J}_{\rm bkg}$ become dominant instead (see the 
red and blue lines in the bottom-left sub-panel of Figure \ref{fig:Fig2} 
for ${\bf J}_{\rm local}$ and ${\bf J}_{\rm bkg}$). The phase of the dipole 
anisotropy turns to the direction of Galactic center. It is noteworthy 
that compared with the traditional diffusion model, the corresponding 
amplitude of CR anisotropy, which is dominated by the background 
${\bf J}_{\rm bkg}$, is naturally suppressed within a SDP model
\cite{2018PhRvD..97f3008G}.

\begin{figure}[!ht]
\includegraphics[width=0.7\textwidth]{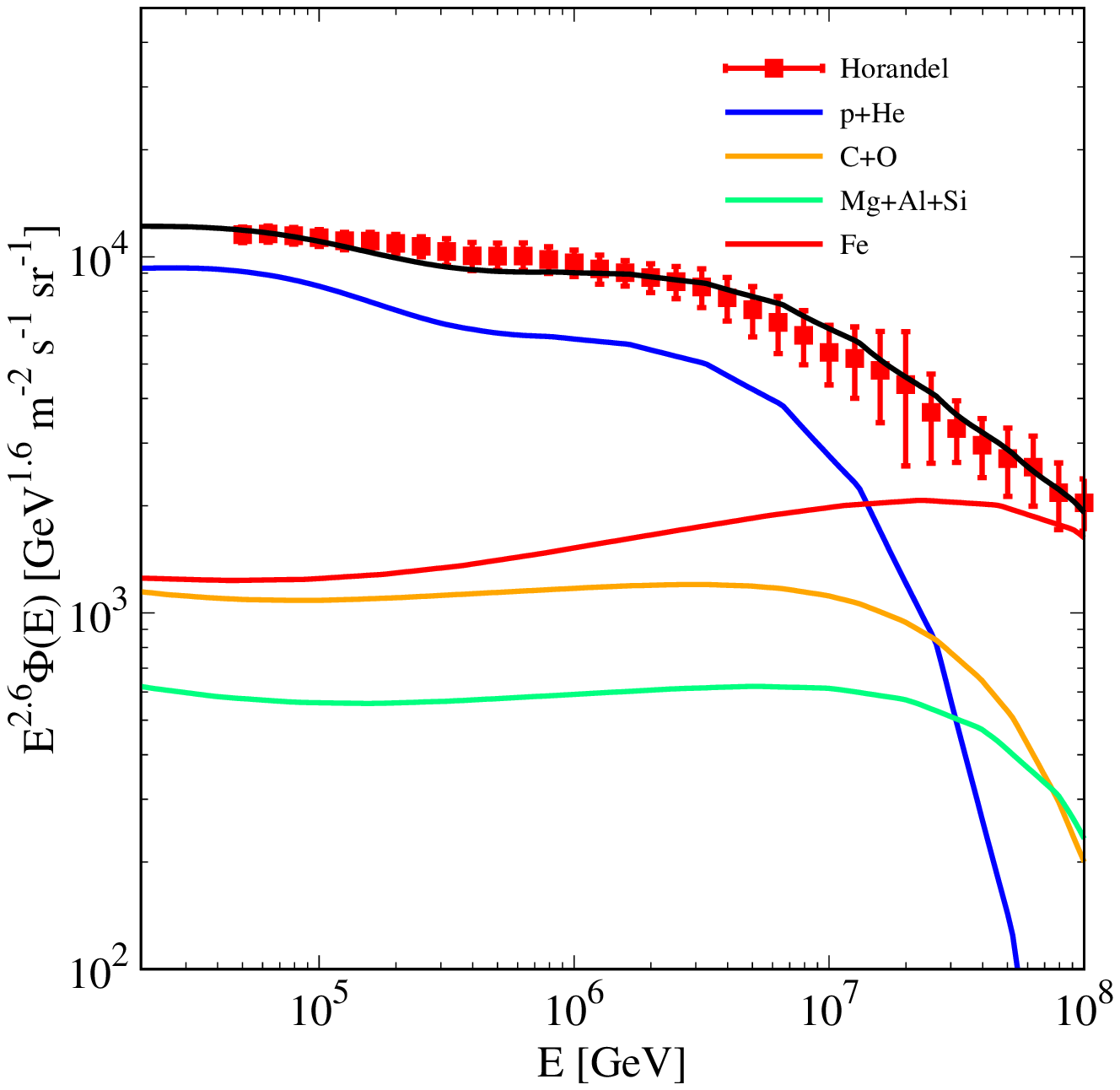}
\caption{The all-particle spectra multiplied by $E^{2.6}$.
The data points are taken from ref.\cite{2003APh....19..193H}.
The solid lines with different colors are the model predictions of different mass groups, and the black solid line is the total contribution.
}
\label{fig:EDFig4}
\end{figure}

As a consistency check, we further calculate the all-particle spectra of 
GCRs, as shown in Figure \ref{fig:EDFig4}. Here we do not consider nuclei 
heavier than helium for the local source. The model prediction is well 
consistent with the observational data \cite{2003APh....19..193H}.

\begin{table*}
\begin{center}
\begin{tabular}{|c|ccc|}
   \hline
Element & Normalization$^\dagger$ & ~~~$\nu$~~~  & ~~~$\mathcal R_{c}$ \\
   \hline
  & $[({\rm m}^2\cdot {\rm sr}\cdot {\rm s}\cdot {\rm GeV})^{-1}]$ & & [PV] \\
     \hline
   p    & $3.75\times 10^{-2}$    & 2.40    &  6.5   \\
   He & $2.4\times 10^{-3}$   & 2.33    &  6.5  \\
   C   & $8\times 10^{-5}$   & 2.35    &  6.5   \\
   O   & $9.8\times 10^{-5}$   & 2.37    &  6.5   \\
   Mg & $1.67\times 10^{-5}$   & 2.33    &  6.5  \\
   Al   & $2.54\times 10^{-6}$   & 2.35    &  6.5  \\
   Si   & $1.43\times 10^{-5}$    & 2.44    &  6.5    \\
   Fe  & $1.37\times 10^{-5}$    & 2.28    &  6.5    \\
\hline
\end{tabular}\\
$^\dagger${The normalization is set at kinetic energy per nucleon $E_k = 100$ GeV/n.}
\end{center}
\caption{Injection parameters of the background sources.}
\label{tab:para_inj}
\end{table*}

\section{Discussion} 

\begin{figure}[!ht]
\includegraphics[width=0.48\textwidth]{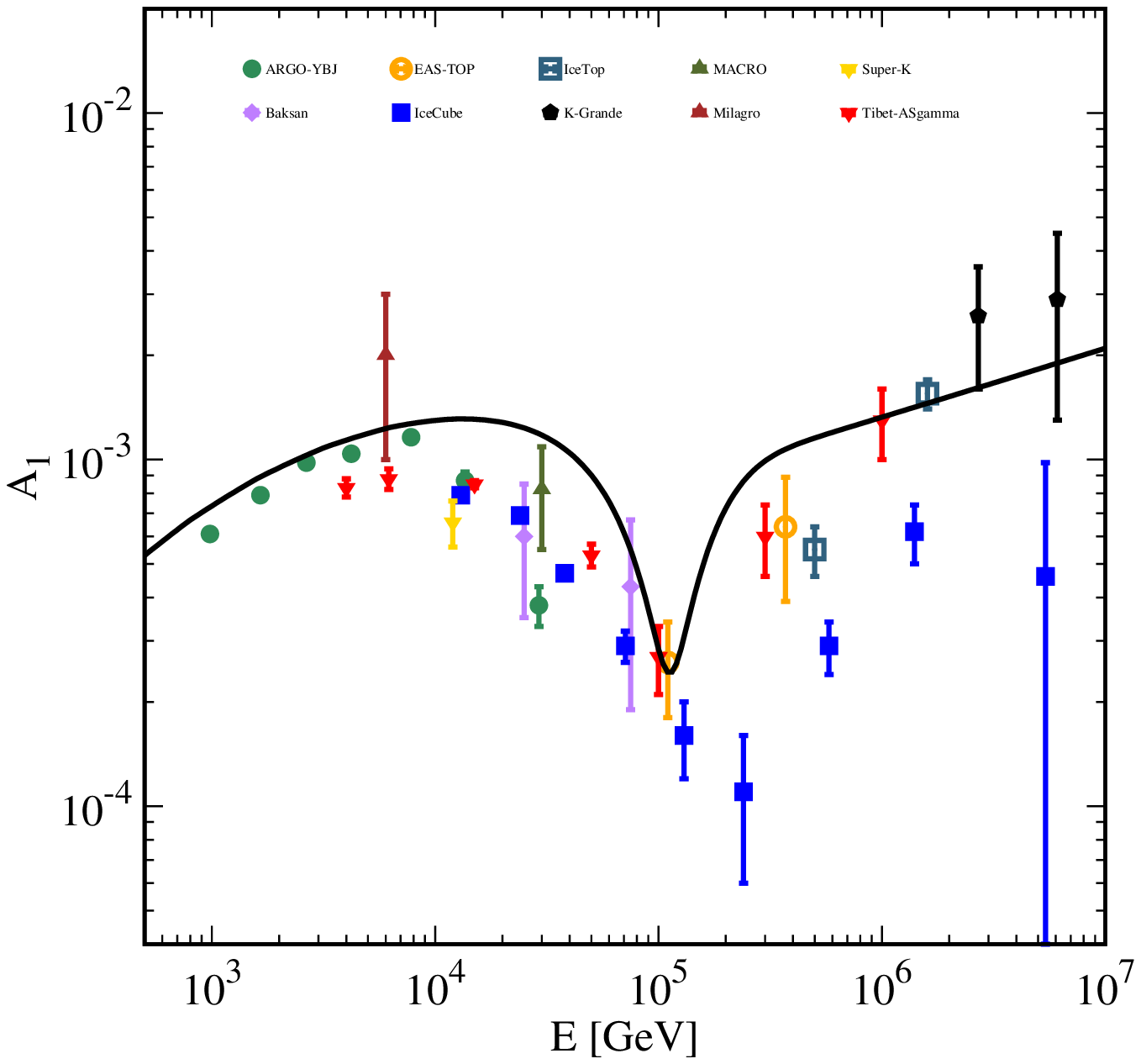}
\includegraphics[width=0.48\textwidth]{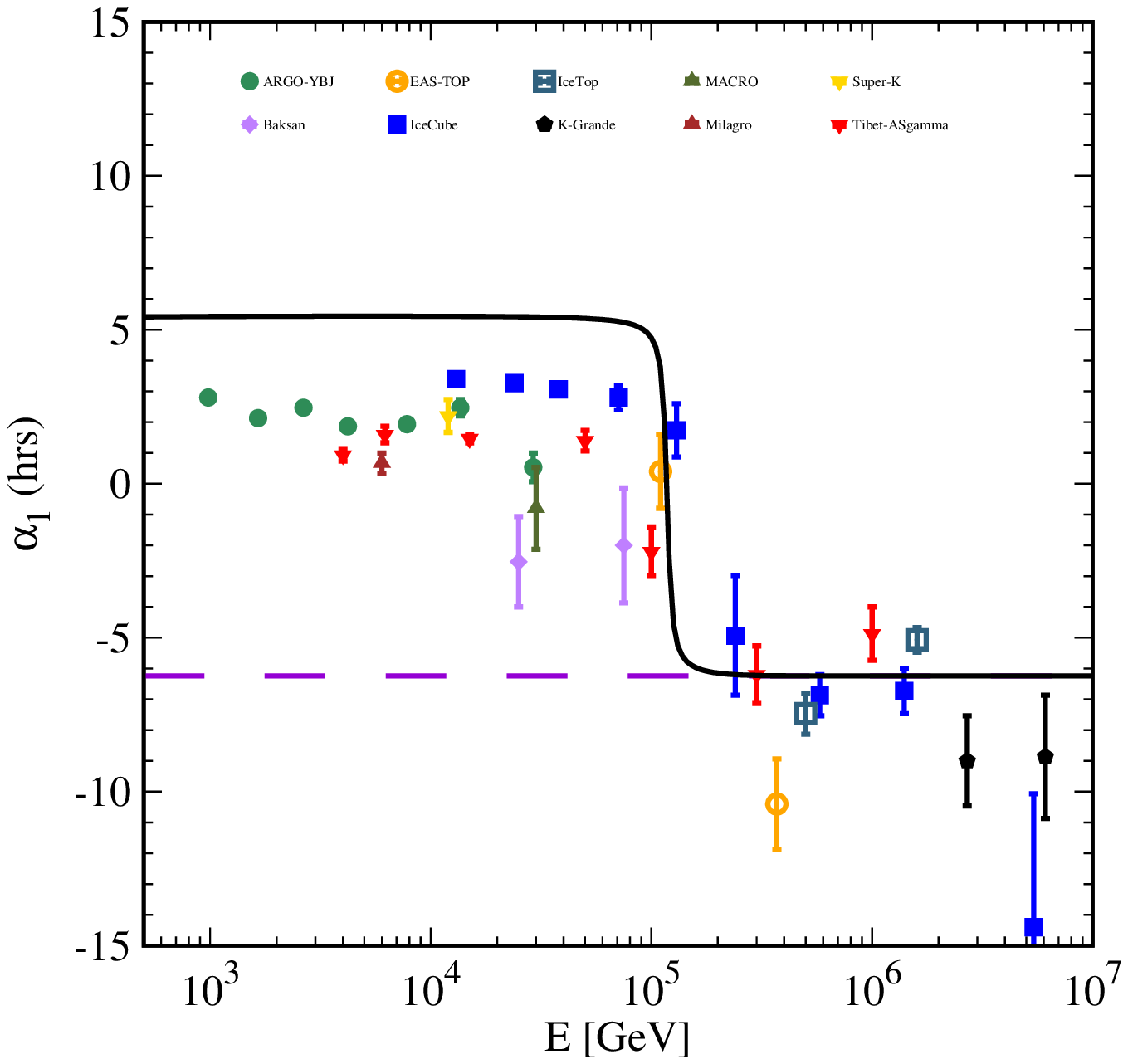}
\caption{Amplitude (left) and phase (right) of the dipole anisotropies
in the SDP scenario together with the contribution from Geminga (at its 
birth place).
}
\label{fig:EDFig2}
\end{figure}

After surveying the catalogues of local SNRs and pulsars, 
we find that the direction close to the Orion association (R.A.$=5^h30^m$,
$\delta=10^{\circ}0'$), which is estimated to be the birthplace of the
Geminga pulsar \cite{1994A&A...281L..41S,2007Ap&SS.308..225F}, is close to
the above required direction. Adopting the source location of the Orion
association (about $330$ pc \cite{1994A&A...281L..41S,2007Ap&SS.308..225F}),
the amplitude and the phase of the anisotropy can be roughly reproduced, 
as shown in Figure \ref{fig:EDFig2}. This result suggests that Geminga is 
probably the dominant source resulting in the spectral and anisotropy 
properties of GCRs from $\sim 1$ to $100$ TeV.



Recent observations in the very-high-energy $\gamma$-ray band by
the High Altitude Water Cherenkov (HAWC) observatory revealed
extended emission around Geminga and another pulsar, which suggested
a slow diffusion of GCR particles in a region of at least a few tens
parsec around these pulsars \cite{2017Sci...358..911A}. Compared with
the diffusion coefficient inferred from the secondary-to-primary ratio
of GCRs \cite{2017PhRvD..95h3007Y}, the HAWC observations suggest
that the diffusion of particles in the Milky Way is
non-uniform \cite{2017PhRvD..96j3013H,2018ApJ...863...30F}.
Therefore the SDP scenario is supported by the HAWC data. Interestingly, 
the modeling of non-uniform diffusion of positrons in light of HAWC 
observations showed that Geminga can be a natural source of the positron 
anomaly \cite{2009Natur.458..607A,2013PhRvL.110n1102A}. Our study further 
indicates that the SNR associated with Geminga could be the source of 
GCR nuclei, which gives rise to the spectral bumps around $10$ TeV of 
the proton and helium spectra and the change of the anisotropy pattern 
around $100$ TeV.

\begin{figure}[!ht]
\includegraphics[width=0.48\textwidth]{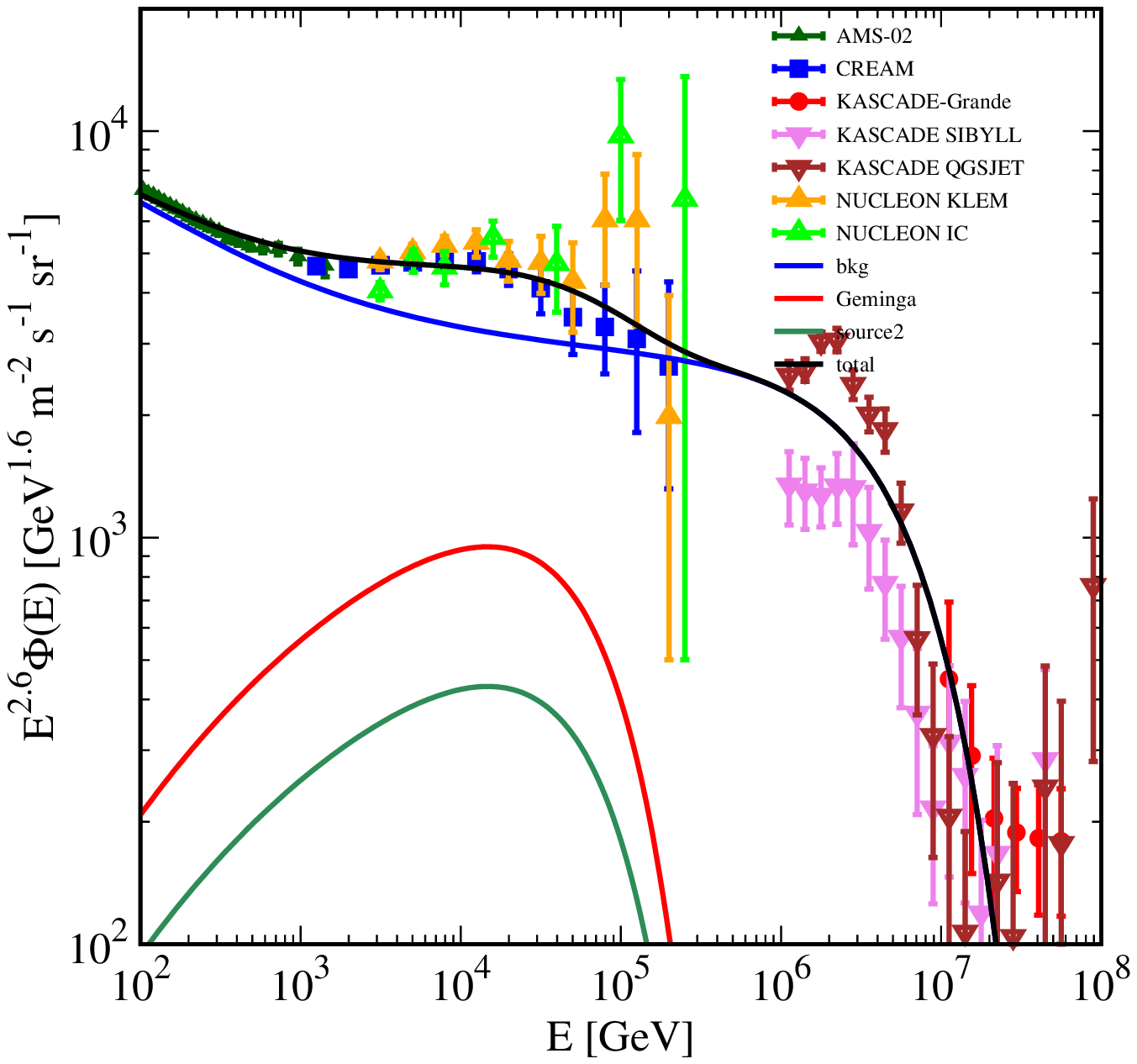}
\includegraphics[width=0.48\textwidth]{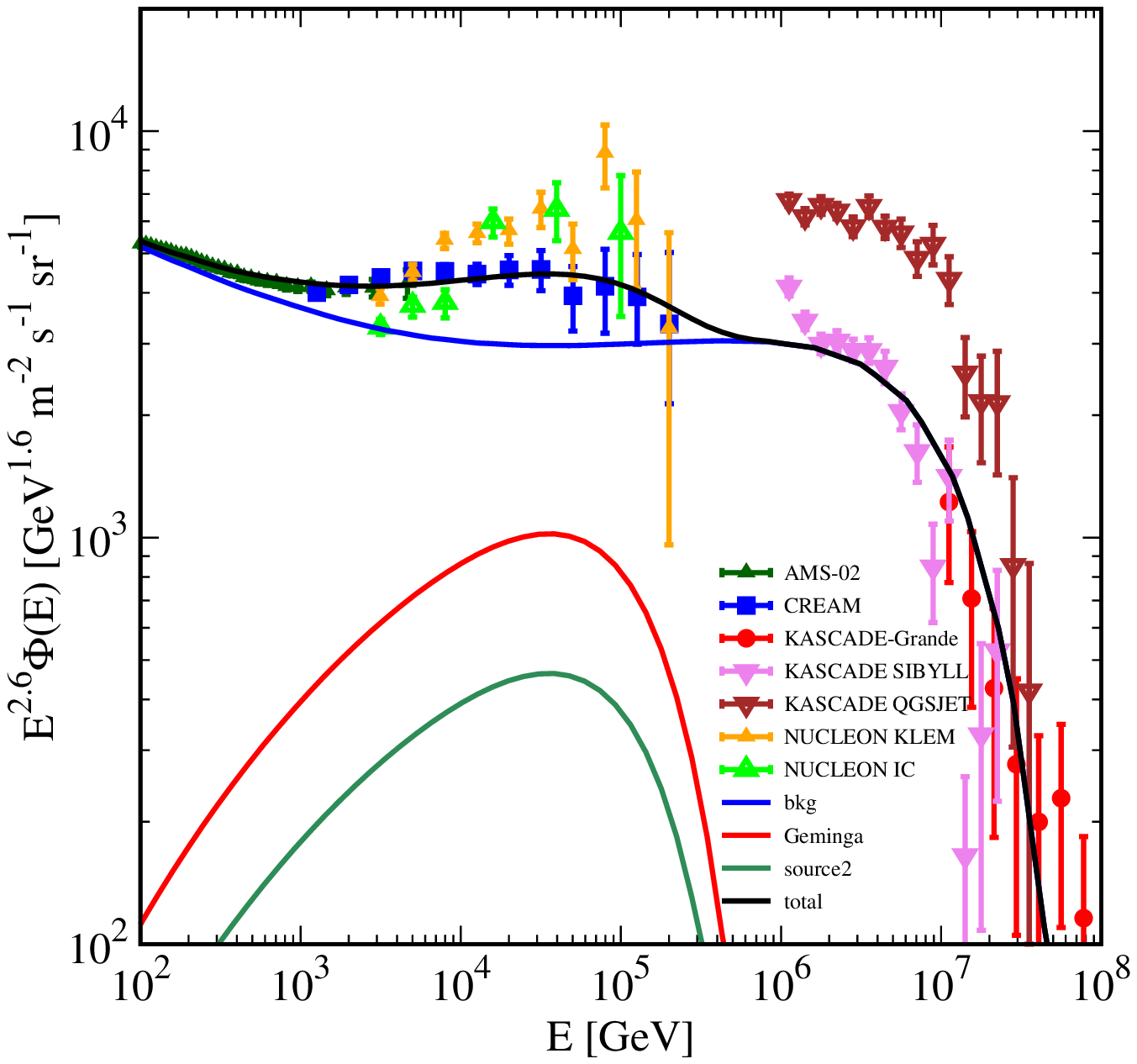}
\includegraphics[width=0.48\textwidth]{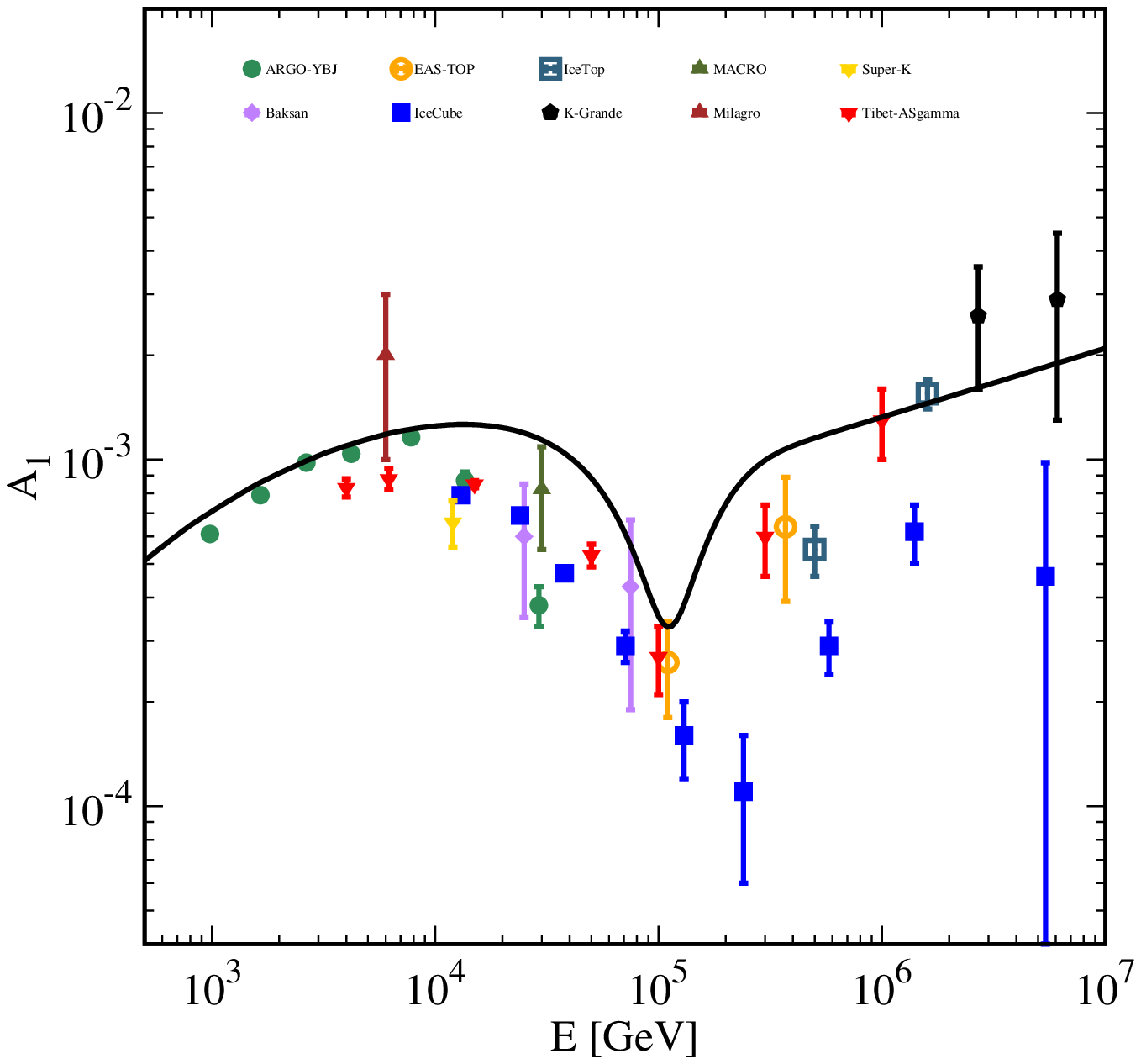}
\includegraphics[width=0.48\textwidth]{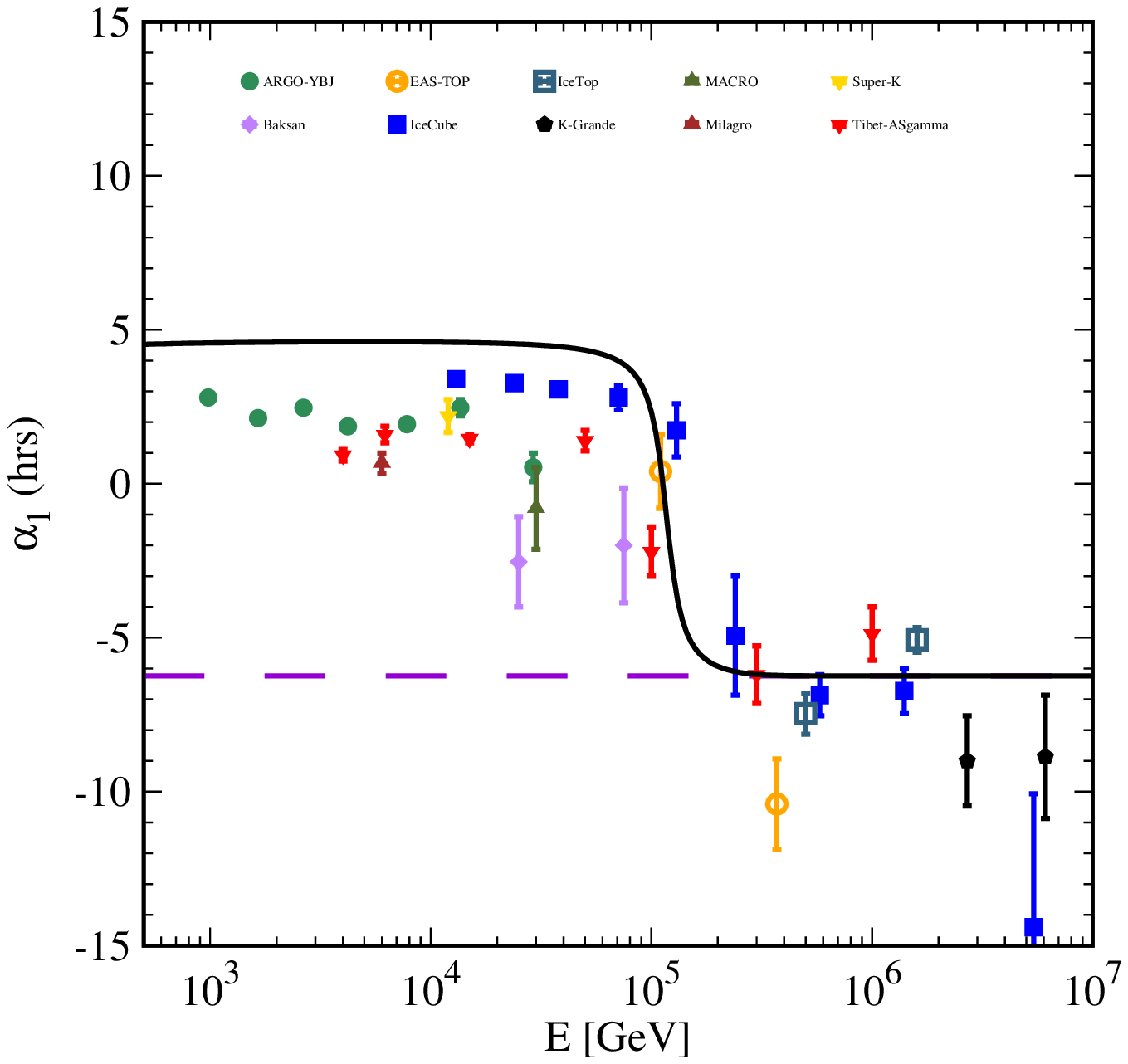}
\caption{Energy spectra of protons (top left), helium nuclei     
(top right), and the amplitude (bottom left) and phase (bottom right)
of the dipole anisotropies. The model consists of a background component 
and two local sources, Geminga (at its birth place) and a hypothetical
source located at (R.A.$=2^h31^m$, $\delta=-8^{\circ}47'$).
}
\label{fig:EDFig3}
\end{figure}

From Figure \ref{fig:EDFig2} we can see that the observational amplitude 
can be quite well reproduced, the phase at the low energy region (below 
100 TeV) is not perfectly consistent with the data. It is possible that 
additional nearby sources other than Geminga also contribute to the 
anisotropies and/or spectra. This scenario should be natural. If SNRs are 
indeed the sources of GCRs, a simple estimate of SNRs in the local vicinity
with proper distances and ages would lead to a number of a few, assuming
a typical rate of Galactic supernovae \cite{2018SCPMA..61j1002Y}. 
We have added one additional source in the direction of 
(R.A.$=2^h31^m$, $\delta=-8^{\circ}47'$), with a distance of 300
pc and an age of $3\times 10^5$ years in the model. We find that 
the fit to the anisotropy phase can be improved with the anisotropy 
amplitude and GCR spectra almost unchanged. The results are shown in 
Figure \ref{fig:EDFig3}.

It was proposed that the anisotropic diffusion due to the large-scale 
magnetic field might result in a projection of the GCR streaming along the
direction of the magnetic field \cite{2010ASTRA...6...49A,
2014Sci...343..988S,2015ApJ...809L..23S,2016PhRvL.117o1103A,
2015PhRvL.114b1101M}, which might account for the low energy ($\lesssim 100$ 
TeV) part of the anisotropies. Therefore, the possible projection effect 
of the anisotropies along the local magnetic field may improve the fit of 
the low energy anisotropy phase of the current model. 
Nevertheless, to what energies the projection effect gets 
to fail and the anisotropies start to reflect the source distribution may 
need further studies in order to properly reproduce the phase change around 
100 TeV energies.

Finally, it is noteworthy that at $E\sim 100$ TeV, the variations of both
the amplitude and phase of anisotropies are very sharp, which can be used 
as an energy calibration for ground-based experiments. Future experiments 
are expected to be able to measure the transition point around $100$ TeV 
accurately.

\section{Summary}

In this work, we propose a two-zone diffusion scenario together with a 
nearby source to explain the energy spectra and anisotropies of GCRs. 
The spectral bumps of GCR protons and helium, reported recently by
CREAM and NUCLEON, can be well fitted by a background component and
a local source component of GCRs. The sum of the streamings of the
background and local source components, can naturally explain the
spectral evolutions of both the amplitude and phase of the dipole
anisotropies. At low energies ($\lesssim100$~TeV), the local source term 
dominates the GCR streaming and determines the low energy anisotropy 
pattern. From the phase of the dipole anisotropy, we propose that the
SNR associated with Geminga may be an important candidate source 
forming the spectral features of GCR spectra and anisotropies. 
For $E\gtrsim100$~TeV, the background component dominates instead, 
and the anisotropy phase points from the Galactic center to
the anti-center, and the amplitude increases with energies again 
following the diffusion law. The SDP scenario, as motivated by the 
HAWC observations of diffuse $\gamma$-ray halos around pulsars, 
suppresses the overall amplitude of the background component.

Our model is quite simple, and well-motivated by up-to-date precise 
observations of GCRs and $\gamma$-rays. In particular, the common energy 
scale appeared in both the monopole (spectra) and dipole (anisotropies) 
can be naturally explained in this model. We link 
the anisotropy spectral evolution with the particle spectra, which show 
the same characteristic energy scale. Importantly, our scenario provides 
a new way to pinpoint the sources of GCRs via spectral features of both 
the fluxes and the anisotropies, which could be applied further to the 
energy range above the knee.

\section*{Acknowledgments}
This work is supported by the National Key Research and Development Program of 
China (No. 2016YFA0400200), the National Natural Science Foundation of China 
(Nos. 11875264, 11635011, 11663006, 11761141001, 11722328, 11851305), and the 
100 Talents program of Chinese Academy of Sciences.

\bibliographystyle{unsrt_update}
\bibliography{ref}

\end{document}